\documentclass[granma]{svjour}
\usepackage{graphicx}
\usepackage{amsmath}
\newcommand{\Sa}{$^{\mathrm{b}}$}
\newcommand{\Sb}{$^{\mathrm{c}}$}
\newcommand{\Sc}{$^{\mathrm{d}}$}
\newcommand{\Sd}{$^{\mathrm{e}}$}
\begin{document}
\title{Heterogeneity and patterning in the quasi-static behavior
of granular materials}
\author{Matthew R. Kuhn}
\offprints{Matthew R. Kuhn}
\institute{%
Dept. of Civil and Env. Engineering\\
School of Engineering\\
University of Portland\\
5000 N. Willamette Blvd., Portland, OR 97203, U.S.A.,\\
kuhn@up.edu,
fax 503-943-7316, tel 503-943-7361}
\date{Received: date / Revised version: date}
\maketitle
\begin{abstract}
Heterogeneity is classified in five categories---topologic, geometric,
kinematic, static, and constitutive---and the first four categories
are investigated in a numerical DEM simulation of
biaxial compression.
The simulation experiments show that
the topology and geometric fabric become more variable during loading.
The measured fluctuations in inter-particle movements are large, they
increase with loading, and they extend to distances of at
least eight particle diameters.
Deformation and rotation heterogeneity are large and are
expressed in spatial patterning.
Stress heterogeneity is moderate throughout loading.
\end{abstract}
\section{Introduction} \label{sec:intro}
Perhaps the most distinguishing characteristic of
granular materials is their internal heterogeneity,
particularly when viewed at the micro-scale of
individual particles or particle clusters.
Granular materials often consist of a wide range of particle
sizes and shapes, and these particles are usually arranged
in an irregular manner.
This geometric and topologic multiformity produce
nonuniform distributions of internal force and deformation,
which are often expressed in spatial and temporal patterning.
In the paper, we catalog the many forms in which heterogeneity
may be manifest, and we provide a classification
scheme for its measurement.
Examples of several forms of heterogeneity are
presented, and certain expressions of their evolution and
spatial patterning are described.
Although the proposed classification scheme applies to both
two- and three-dimensional (2D and 3D) granular materials, to
particles of arbitrary shape and composition, to both sparse and dense
packings, and to both dynamic and quasi-static deformations,
the paper illustrates the classification within a two-dimensional
framework and with a 2D example of the quasi-static deformation
of a dense disk assembly.
\par
In Section~\ref{sec:classification} we consider
a classification scheme
for heterogeneity and the various forms in which it can be expressed
and measured.
Section~\ref{sec:methods} describes the simulation methods that
are used to explore several forms of heterogeneity.
In the same section, we also consider a means of
measuring the heterogeneity of vector and tensor objects.
Section~\ref{sec:results} presents experimental results and characterizes
several types of heterogeneity and their evolution during
biaxial compression.
\section{Classifying heterogeneity} \label{sec:classification}
Table~\ref{table:class1} gives a classification of material characteristics
that can manifest heterogeneity in granular materials.
\begin{table}
  \centering
  \caption{Heterogeneity categories and references to experimental studies}
  \label{table:class1}
  \begin{tabular}{ll}
\hline\noalign{\smallskip}
Category$^{\mathrm{a}}$ & References\\
\hline\noalign{\smallskip}
Topologic (\ref{sec:topology})&\\
  \quad{Coordination number}
        &\cite{Smith:1929a,Oda:1977a,Lemaitre:1993a,Yang:2000a}\Sa,
         \cite{Chapuis:1976a,Oda:1982a,Bathurst:1990a,Chen:1990a,Kuhn:1999a,Cambou:2000a,Thornton:2000a}\Sd \\

  \quad{Valence}
	&\cite{Annic:1993a,Kuhn:1999a}\Sa \\
	
Geometric&\\
  \quad{Grain size} 
        &              \\
	
  \quad{Grain shape} 
        &              \\
	
  \quad{Grain orientation}
        &\cite{Oda:1972b,Oda:1993a,Ng:2001a},
         \cite{Oda:1972a,Rothenburg:1992b}\Sa \\
	
  \quad{Void ratio}
        &\cite{Matsushima:2001a},
         \cite{Lemaitre:1993a,Mogami:1966a,Hinrichsen:1990a}\Sa,\\
        & \quad
         \cite{Hinrichsen:1990a}\Sb,
         \cite{Bardet:1991a,Desrues:1996a,Iwashita:1998a,Sture:1998a,Frost:2000a}\Sc \\
	
  \quad{Void size/shape}
	&{\cite{Tsuchikura:2001a}\Sa} \\
	
  \quad Branch vector 
        &\cite{Oda:1972b,Ng:2001a,Rothenburg:1993a,Wang:2001a},
         \cite{Rothenburg:1992a}\Sa,\\
  \quad\quad orientation
        &\quad\cite{Oda:1982a,Oda:1993a,Calvetti:1997a,Daudon:1997a}\Sd \\
	
  \quad{Branch vector length}
	&\cite{Yang:2000a},\cite{Kruyt:2001a}\Sb \\
	
  \quad{Fabric tensor} (\ref{sec:fabric})
	&\cite{Oda:1980a,Satake:1982a},
         \cite{Chen:1990a,Bathurst:1990a,Thornton:2000a}\Sd \\
	
  \quad{Loop tensor}
        &\cite{Kuhn:1999a,Konishi:1988a},
         \cite{Tsuchikura:2001a}\Sa \\
	
Kinematic&\\
  \quad{Particle movement}
        &\cite{Masson:2001a},
         \cite{Bagi:1993a}\Sa,
         \cite{Cambou:2001a,Kruyt:2002a}\Sb,\\
        & \quad
         \cite{Kuhn:1999a,Williams:1997b,Misra:1997a}\Sc \\
	
  \quad{Inter-particle motion} (\ref{sec:move})
	&\cite{Bathurst:1988a,Kruyt:2001a},
         \cite{Kuhn:1999a,Bardet:1994a}\Sc, \\

  \quad{Particle rotation} (\ref{sec:rotate})
        &\cite{Oda:1982a,Calvetti:1997a,Daudon:1997a,Bagi:1993a,Misra:1997a}\Sa,\\
        &\quad
         \cite{Kuhn:1999a,Matsushima:2001a,Iwashita:1998a,Masson:2001a}\Sc \\
	
  \quad{Deformation} (\ref{sec:deform})
	&\cite{Bagi:1993a}\Sa,
         \cite{Bardet:1991a,Calvetti:1997a,Kuhn:1999a,Dedecker:2000a}\Sc \\
	
Static&\\
  \quad{Contact force}
        &\cite{Masson:2001a,Matsuoka:1974a,Thornton:1990a,Antony:2000a},
         \cite{Yang:2000a,Cundall:1989b,Gherbi:1993a,Radjai:1997a,Radjai:1998a,Mueth:1998a,Oger:1998a,Calvetti:1999a,Kruyt:2002b}\Sa,\\
        &\quad
         \cite{Oger:1998a,Kruyt:2002b}\Sa,
         \cite{deJosselin:1969a,Thornton:1990a,Iwashita:1998a,Masson:2001a}\Sc,\\
        &\quad
         \cite{Cambou:2000a}\Sd \\
	
  \quad{Force potential}
	&\cite{Kruyt:2002a}\Sb \\
	
  \quad{Stress} (\ref{sec:stress})
	&\cite{Auvinet:1992a,Bacconnet:1992a}\Sa \\
	
Constitutive&\cite{Gaspar:2001b,Gaspar:2002a}\Sa\\
\hline\noalign{\smallskip}
\multicolumn{2}{l}{%
$^{\mathrm{a}}$\ Section numbers (*.*) refer to the experimental results}\\
\multicolumn{2}{l}{%
\ \ in this paper}\\

\multicolumn{2}{l}{%
\Sa\ Includes statistical measures, e.g. including standard}\\
\multicolumn{2}{l}{%
\ \ deviations or histograms}\\
\multicolumn{2}{l}{%
\Sb\ Includes spatial correlations}\\
\multicolumn{2}{l}{%
\Sc\ Includes spatial patternings}\\
\multicolumn{2}{l}{%
\Sd\ Includes rates of changes or temporal patterning}\\
\hline\noalign{\smallskip}
\end{tabular}

\end{table}
The table references sample experimental studies in which these 
characteristics have been measured, although the short lists of references
are far from exhaustive.
The characteristics in the table are organized within a hierarchy of 
heterogeneity categories:  topologic,
geometric, kinematic, static, and constitutive.
These categories are described in a general manner in the next paragraph.
Table~\ref{table:class2} presents a short list of informational
forms that can be used for describing each characteristic.
\begin{table}
\caption{Analyses of heterogeneity}
\label{table:class2}
  \centering
\begin{tabular}{ll}
\hline\noalign{\smallskip}
Informational forms & Examples\\
\noalign{\smallskip}\hline\noalign{\smallskip}
Central tendency & Mean, median, modes\\
Dispersion & Standard deviation, \\
           & variance, \\
           & coefficient of variation,\\
           & histograms,\\
           & probability and cumulative \\
           & \quad distributions,\\
           & quartile plots\\
Spatial correlation & n-point correlations,\\
                    & correlation lengths\\
Temporal correlation & Rates of change \\
Spatial and temporal & Spatial plots,\\
\quad patterning     & time series analyses,\\
                     & spatial domain transforms\\
\noalign{\smallskip}\hline
\end{tabular}
\end{table}
The arrangement of the forms in Table~\ref{table:class2} 
reflects their complexity and the
usual historical order in which measurements have been proposed
and collected.
The simplest form of information is some measure of central tendency:
a mean, median, or modal value.
Heterogeneity implies diversity and fluctuation, and this 
dispersion in measured
values can be expressed as a variance or standard deviation,
with standard graphical means such as histograms, or by
fitting experimental results to an appropriate probability distribution.
Of greater complexity are measurements of temporal correlation
(e.g. rates of change) and spatial correlation.
The most complex data analyses can also disclose the spatial and
temporal patterning of heterogeneity.
The paper presents data on the six characteristics
that are accompanied by section numbers in Table~\ref{table:class1}, and 
these characteristics are explored with a range of the
informational forms that are given in
Table~\ref{table:class2}.
\par
Table~\ref{table:class1} begins with topologic characteristics,
which concern the arrangement of the particles and their contacts,
but without reference to their position, size, or orientation.
This information can be expressed as a \emph{particle graph}
for both 2D and 3D assembles, which
gives the topologic connectivity of the particles in a packing,
as explained in~\cite{Satake:1993b}.
The paper presents data on the variation in local topology and
its evolution during loading.
A discrete metric is also proposed as a means of tracking 
inter-particle processes between distant particles.
Geometric information includes the additional descriptors of
length, shape, and angle, which relate to the positional arrangements,
orientations, and sizes of particles.
Together, topology and geometry describe the \emph{fabric} of
a granular assembly.
The paper characterizes the evolution of one form of heterogeneity in
this fabric.  
Kinematic information (Table~\ref{table:class1}) concerns
the movements and rotations of particles, 
relative inter-particle movements, 
and the local deformations within small groups
of particles.
The paper gives examples of heterogeneous movements and
deformations, the spatial correlation of inter-particle movements, and 
the patterning of local rotations and deformations.
Static (or statical) information 
(Table~\ref{table:class1})
involves the transmission of
force and stress within a material, and the paper depicts the local
diversity of stress and its evolution during
loading.
Table~\ref{table:class1} also includes the category of \emph{constitutive}
heterogeneity (or, perhaps, mechanical heterogeneity), which would
involve the diversity in local material stiffness.
Except for simple two-particle models that rely on uniform
strain assumptions, there is, as of yet, no consistent vocabulary or
experimental methods for measuring and characterizing
this form of heterogeneity.
The reader is referred to the recent work of 
Gaspar and Koenders~\cite{Gaspar:2001b} and Gaspar~\cite{Gaspar:2002a},
which may
provide a needed framework for characterizing constitutive
heterogeneity.
\par
As a simple example of the classification scheme in Table~\ref{table:class2},
we could consider the diversity of grain size in a granular material.
Methods for measuring and describing particle size, such
as sieving methods, are standardized and widely applied, so
that references to these methods are excluded from Table~\ref{table:class2}.
These methods can readily depict a representative (central)
grain size as well as the dispersion of sizes.
Certain processes, such as shearing and compression, can cause
particle breakage, which could be measured with temporal correlations of
the size distribution.
Processes that promote size segregation could be studied with
methods that reveal the spatial correlation of size.
Size segregation usually leads to a spatial patterning of the local
size distribution, and processes that produce a periodic recurrence in
such patterning would lead to both spatial and temporal patterning.
\section{Methods and notation} \label{sec:methods}
A conventional implementation of the Discrete Element Method (DEM)
was used to simulate the quasi-static behavior of a large 2D 
granular assembly and to illustrate different manifestations
of internal heterogeneity and their evolution.
\subsection{Simulation methods}
The study employs a square assembly containing 10,816 circular
disks of multiple diameters.
The disk sizes are randomly distributed over
a fairly small range of between 0.56$\overline{D}$ and 1.7$\overline{D}$,
where $\overline{D}$ is the mean particle diameter.
The material was created by slowly and isotropically compacting
a sparse arrangement of particles, during which friction between particle
pairs was disallowed
(friction was later restored for biaxial compression tests).
This compaction technique produced a 
material that was dense, random, and isotropic,
at least when viewed at a macro-scale.
The average initial void ratio was 0.1715 (solid fraction of $0.854$),
the average coordination number was 3.95, and the average overlap between neighboring particles was
about 9$\times$10$^{-4}$ of $\overline{D}$.
The assembly was surround by periodic boundaries, a choice that would
eliminate the topologic and geometric nonuniformity that
might otherwise occur in the vicinity of rigid platens or assembly
corners.
The initial height and width of the assembly were each about 
$102\overline{D}$.
\par
All examples of heterogeneity were collected from a single loading test
of biaxial compression.
The height of the assembly was reduced at a constant rate of compressive
strain ($\dot{\varepsilon}_{22}<0$), while maintaining a constant average
horizontal stress ($\dot{\sigma}_{11}=0$).
About 200,000 time steps were required to reach
the final vertical strain, $\overline{\varepsilon}_{22}$, of $-0.01$, 
and at this 
rate of loading, the average imbalance of
force on a particle was less than 1$\times$10$^{-4}$
times the average contact force.
\par
During biaxial compression,
a simple force mechanism was employed between contacting particles.
Linear normal and tangential contact springs were assigned equal
stiffnesses ($k_{\mathrm{n}}=k_{\mathrm{t}}$), 
and slipping between particles would occur whenever
the contact friction coefficient of 0.50 was attained.
\par
\begin{figure}
  \centering
  \includegraphics{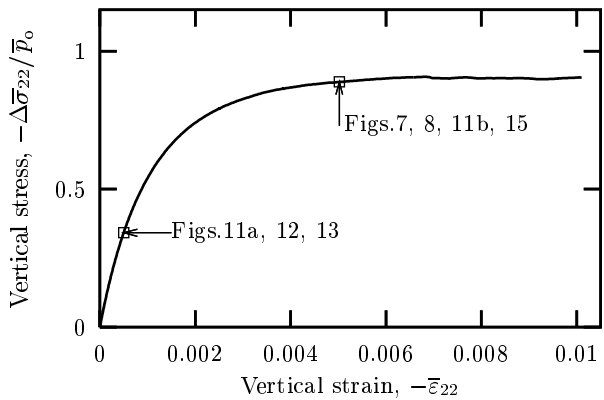}
\caption{Evolution of the average compressive stress within the assembly
of 10,816 circular disks during biaxial compression.}
\label{fig:crs_q}
\end{figure}

The average, macro-scale mechanical behavior is shown in
Fig.~\ref{fig:crs_q}, which gives the dimensionless compressive stress
\mbox{$\Delta\overline{\sigma}_{22}/\overline{p}_{\mathrm{o}}$},
where $\overline{p}_{\mathrm{o}}$ is the initial mean stress,
$\overline{p}_{\mathrm{o}}=(\overline{\sigma}_{11}+\overline{\sigma}_{22})/2$.
This initial mean stress was about 5$\times$10$^{-4}$ times the normal
contact stiffness,~$k_{\mathrm{n}}$.
\par
The rates of several micro-quantities 
(position, force, orientation, etc.)
were periodically measured during
the loading.
These rates were calculated by first collecting
the assembly's status at two instants that were separated by
100 time steps, and the difference in these states was then used
to compute the rates.
Because time is used in quasi-static DEM simulations
as simply a means of ordering or parameterizing events,
the rates of micro-quantities will usually be expressed
in a dimensionless form by dividing by an average, 
macro-scale rate (average stress rate, average strain rate, etc.).
\subsection{Notation}\label{sec:notation}
Vectors and tensors are represented by bold Roman letters,
lower and upper case respectively.
Their inner products are computed as
\begin{equation} \label{eq:innerp}
\mathbf{a} \cdot \mathbf{b} = a_{p}a_{p}, \quad
\mathbf{A} \cdot \mathbf{B} = A_{pq}B_{pq},
\end{equation}
with the associated norms
\begin{equation} \label{eq:norm}
|\mathbf{a}| = (\mathbf{a} \cdot \mathbf{a})^{1/2}, \quad
|\mathbf{A}| = (\mathbf{A} \cdot \mathbf{A})^{1/2}.
\end{equation}
A juxtaposed tensor and vector will represent the
conventional product
\begin{equation}
\mathbf{A} \mathbf{b} = A_{pq}b_{q},
\end{equation}
and juxtaposed tensors represent the product
\begin{equation}
\mathbf{A} \mathbf{B} = A_{pr}B_{qr}.
\end{equation}
Various quantities are measured at both micro and macro scales
so that the variability of the micro-scale measurements
can be deduced.
A macro-level, assembly average is indicated with
an overline ($\overline{\mathbf{L}}$, $\overline{\sigma}_{22}$, 
$\overline{p}_{\mathrm{o}}$, $\overline{q}$);
whereas local, micro-level quantities appear with superscripts
($\mathbf{L}^{i}$, $\boldsymbol{\sigma}^{k}$, $\widehat{\mathbf{v}}^{j}$,
$p^{k}$, Table~\ref{table:superscripts}).
\begin{table}
\caption{Superscript notation}
\label{table:superscripts}
  \centering
\begin{tabular}{cl}
\hline\noalign{\smallskip}
Index & Usage \\
\noalign{\smallskip}\hline\noalign{\smallskip}
$i$ & A polygonal void cell having $m^{i}$ edges and \\
    & vertices.  An $m$-tuple of particles or contacts,\\
    & $i=(k_{1},k_{2},\ldots,k_{m^{i}})$ or 
      $i=(j_{1},j_{2},\ldots,j_{m^{i}})$\\
$j$ & A contacting pair of particles $(k_{1},k_{2})$\\
$k$ & A single particle\\
\noalign{\smallskip}\hline
\end{tabular}
\end{table}
The ``$k$'' superscript is used with quantities that can be measured
within a single particle or its immediate vicinity;
the ``$i$'' superscript is assigned to quantities that are
measured within a single void cell (the dual of particles);
and the ``$j$'' superscript is used for quantities associated
with a pair of particles or a pair of void cells
(e.g. contacts, contact forces, branch vectors, 
and inter-particle velocities).
No contractive summation is implied with superscripts,
e.g. $a^{j}b^{j}$.
\par
The non-uniformity of scalar, vector, and tensor quantities is
considered in the paper.
A consistent notation is used to express the conformity (or diversity)
of a local quantity $\mathbf{a}^{\mathrm{local}}$
with respect to
the corresponding assembly average $\overline{\mathbf{a}}$.
The pair $\mathbf{a}^{\mathrm{local}}$ and $\overline{\mathbf{a}}$
may be scalars, vectors, or tensors.
Three dimensionless scalars measure the \emph{participation}
of $\mathbf{a}^{\mathrm{local}}$ 
($= \mathbf{a}^{\mathrm{local}} \!\parallel \overline{\mathbf{a}}$) 
in the assembly-average $\overline{\mathbf{a}}$;
the \emph{non-conformity} of $\mathbf{a}^{\mathrm{local}}$
($= \mathbf{a}^{\mathrm{local}} \!\perp \overline{\mathbf{a}}$);
and the \emph{alignment} of $\mathbf{a}^{\mathrm{local}}$
($= \mathbf{a}^{\mathrm{local}} \!\circ \overline{\mathbf{a}}$)
with respect to the assembly-average~$\overline{\mathbf{a}}$:
\begin{align}
\mathbf{a}^{\mathrm{local}} \!\parallel \overline{\mathbf{a}} &=
  \frac{1}{|\overline{\mathbf{a}}|^{2}}
  \left( \mathbf{a}^{\mathrm{local}} \cdot \overline{\mathbf{a}}\right)
  \label{eq:parallel}\\
\mathbf{a}^{\mathrm{local}} \!\perp \overline{\mathbf{a}}     &=
  \frac{1}{|\overline{\mathbf{a}}|}
  \left| \mathbf{a}^{\mathrm{local}} - 
         (\mathbf{a}^{\mathrm{local}} \parallel \overline{\mathbf{a}})
         \overline{\mathbf{a}}
  \right|
  \label{eq:perp}\\
\mathbf{a}^{\mathrm{local}} \!\circ \overline{\mathbf{a}}     &=
  \frac{1}{|\mathbf{a}^{\mathrm{local}}|\,|\overline{\mathbf{a}}|}
  \left( \mathbf{a}^{\mathrm{local}} \cdot \overline{\mathbf{a}}\right)
  \label{eq:circ}
\end{align}
The participation and non-conformity in Eqs.~\ref{eq:parallel} 
and~\ref{eq:perp} are the 
dimensionless magnitudes of $\mathbf{a}^{\mathrm{local}}$
in directions parallel and perpendicular to $\overline{\mathbf{a}}$,
and relative to the length of $\overline{\mathbf{a}}$.
The alignment $\mathbf{a}^{\mathrm{local}} \!\circ \overline{\mathbf{a}}$
is the cosine of the angle separating
$\mathbf{a}^{\mathrm{local}}$ and $\overline{\mathbf{a}}$.
These quantities are unambiguous when $\mathbf{a}$ is a
vector or tensor.
If $\mathbf{a}$ is a scalar,
then $\mathbf{a}^{\mathrm{local}} \!\parallel \overline{\mathbf{a}}$
is simply the quotient $a^{\mathrm{local}}/\,\overline{a}$;
$\mathbf{a}^{\mathrm{local}} \!\!\perp \overline{\mathbf{a}}$
is zero;
and $\mathbf{a}^{\mathrm{local}} \!\circ \overline{\mathbf{a}}$
is sgn($a^{\mathrm{local}},\,\,\overline{a}$).
By reducing vector and tensor objects to the scalars in
Eqs.~(\ref{eq:parallel}--\ref{eq:circ}),
we can compute conventional statistical measures such as the
mean, standard deviation, and coefficient of variation.
These measures will be represented with the notation
$\mathsf{Mean}(\cdot)$, $\mathsf{Std}(\cdot)$, 
and $\mathsf{Cov}(\cdot)$, 
where the coefficient of variation
$\mathsf{Cov}(\cdot) = \mathsf{Std}(\cdot) / \mathsf{Mean}(\cdot)$.
\par
As an example with vector quantities $\mathbf{a}$, we can consider
two different sets of two-dimensional vectors $\mathbf{a}^{\mathrm{local}}$,
and this example can serve as a reference case for comparing
the results given later in the paper.
In both sets, the vectors $\mathbf{a}^{\mathrm{local}}$ all have
unit length.
In the first set, the vectors 
$\mathbf{a}^{\mathrm{local}}$ have a uniform direction that is aligned with
the reference vector $\overline{\mathbf{a}}$;
but in the second set, the vectors $\mathbf{a}^{\mathrm{local}}$
have uniformly random directions.
In the example, the reference vector $\overline{\mathbf{a}}$
is also assumed to have unit length.
The four statistical measures 
$\mathsf{Mean}(\mathbf{a}^{\mathrm{local}} \!\parallel \overline{\mathbf{a}})$,
$\mathsf{Std}(\mathbf{a}^{\mathrm{local}} \!\parallel \overline{\mathbf{a}})$,
$\mathsf{Mean}(\mathbf{a}^{\mathrm{local}} \!\perp \overline{\mathbf{a}})$,
and $\mathsf{Mean}(\mathbf{a}^{\mathrm{local}} \circ \overline{\mathbf{a}})$
are used in the paper as indicators of local non-conformity and
heterogeneity, and their values
for this simple example are
summarized in Table~\ref{table:values}.
\begin{table}
\caption{Statistics of uniform and random vector sets}
\label{table:values}
  \centering
\begin{tabular}{lcc}
\hline\noalign{\smallskip}
&\multicolumn{2}{c}{Vectors $\mathbf{a}^{\mathrm{local}}$}\\
& Uniform,& \\
Measure & aligned & Random \\
\noalign{\smallskip}\hline\noalign{\smallskip}
$\mathsf{Mean}(\mathbf{a}^{\mathrm{local}} \parallel \overline{\mathbf{a}})$ &
  1 & 0 \\
$\mathsf{Std}(\mathbf{a}^{\mathrm{local}} \parallel \overline{\mathbf{a}})$ &
  0 & $1/2$ \\
$\mathsf{Mean}(\mathbf{a}^{\mathrm{local}} \perp \overline{\mathbf{a}})$ &
  0 & $2/\pi$ \\
$\mathsf{Mean}(\mathbf{a}^{\mathrm{local}} \circ \overline{\mathbf{a}})$ &
  1 & 0 \\
\noalign{\smallskip}\hline
\end{tabular}
\end{table}
In the simulated biaxial loading of 10,816 circular disks, certain local
vector and tensor quantities are found to have measured
values of
$\mathsf{Std}(\mathbf{a}^{\mathrm{local}} \!\parallel \overline{\mathbf{a}})$
and $\mathsf{Mean}(\mathbf{a}^{\mathrm{local}} \!\perp \overline{\mathbf{a}})$
that greatly exceed those of the random set, as given in the 
final column of Table~\ref{table:values}.
These large values are due to variations in the magnitudes of the
local quantities as well as in their directions.
\section{Heterogeneity measurements} \label{sec:results}
The experimental results are analyzed for indications of four
categories of heterogeneity:
topologic, geometric (fabric), kinematic, and static.
\subsection{Topologic heterogeneity} \label{sec:topology}
\begin{figure}
  \centering
  \includegraphics{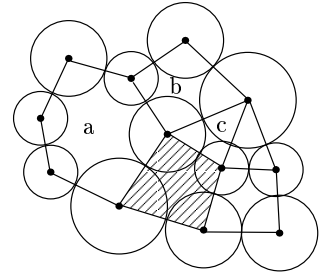}
\caption{Particle graph of a 2D granular assembly.  A single
void cell is shaded.  The void cells labeled~a, b, and~c have
valences of~6, 4, and~3 respectively.}
\label{fig:graph}
\end{figure}

In a 2D setting, the topology of an assembly can be
described by the \emph{particle graph} of its particles
(the graph vertices) and their contacts (the graph edges)~\cite{Satake:1993b}.
The particle graph is associated with the Voronoi-Dirichlet
tessellation of a 2D region, except that the particle
graph admits only the real contacts as graph edges.
The faces of the planar graph are polygonal void cells, which are
enclosed by the circuits of contacting particles
(an example void cell is shaded in Fig.~\ref{fig:graph}).
For this topologic description of a 2D granular material,
the simplest local topologic measures are the local
coordination number $n^{k}$ and the local valence $m^{i}$, defined
as the number of contacts of a single particle $k$, and the number
of edges of a single void cell $i$
(see Fig.~\ref{fig:graph} for examples of valence).
Because gravity is absent in the current simulations,
some particles will be unattached and, hence, excluded from the
particle graph.
The effective average coordination number $\overline{n}_{\mathrm{eff}}$ 
of the attached particles will be somewhat larger than the coordination
number $\overline{n}$ that includes both attached and unattached 
particles~\cite{Kuhn:1999a,Thornton:2000a}.
Dense assemblies have large coordination numbers and small valences,
but during biaxial compression, the average effective
coordination number is reduced, while the average valence
increases~\cite{Kuhn:1999a,Thornton:2000a}.
In the simulation of biaxial compression, $\overline{n}_{\mathrm{eff}}$
is reduced from 4.14 in the initial particle arrangement to a value
of 3.50 at the final compressive strain, $\overline{\varepsilon}_{22}=-0.01$.
The average valence $\overline{m}$ increases from 3.87 to 4.66.
\par
\begin{figure}
  \centering
  \includegraphics{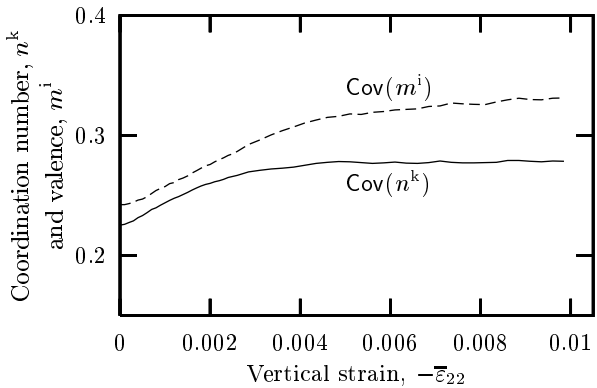}
\caption{The evolution of two measures of topologic heterogeneity 
during biaxial compression: 
the coefficients of variation ($\mathsf{Cof}$) 
of the local coordination number ($n^{k}$)
and local valence ($m^{i}$).}
\label{fig:topology}
\end{figure}

A simple measure of topologic nonuniformity is the dispersion in
the local values of $n^{k}$ and $m^{i}$.
Figure~\ref{fig:topology} shows the evolution of the coefficients
of variation of these two local topologic measures.
Together, the results indicate an increase in topologic heterogeneity during
loading.
The large increase in the dispersion of local valence, 
as expressed by the coefficient of variation $\mathsf{Cov}(m^{i})$,
is consistent with the results of 
Tsuchikura and Satake~\cite{Tsuchikura:2001a},
who have shown that the sizes of void cells become more diverse 
during biaxial compression.
The increase in the coefficient of variation of the local coordination
number, $\mathsf{Cov}(n^{i}) = \mathsf{Std}(n^{i}) / \mathsf{Mean}(n^{i})$,
is due, in part, to a reduction in the mean coordination number.
\subsection{Geometric heterogeneity}\label{sec:fabric}
Geometric characteristics of granular materials are listed
in Table~\ref{table:class1},
and numerous studies have shown how the assembly averages of these
characteristics evolve during loading.
Fewer studies indicate how the internal diversity 
of these characteristics changes with loading.
Tsuchikura and Satake~\cite{Tsuchikura:2001a} have developed
methods for examining the diversity of local fabric
in a 2D granular material and found that void cells become more
elongated during loading, but that the variation in elongation
remains fairly uniform.
To study this form of fabric anisotropy, they
propose a method for computing the magnitude of the anisotropy of a general
second order symmetric tensor $\mathbf{T}$ by considering
its deviatoric part $\mathbf{T}'$.
The self-product of $\mathbf{T}'$ yields a scalar measure $\beta$
of anisotropy:
\begin{equation}\label{eq:beta}
\mathbf{T}' \mathbf{T}' = \beta^2 \,\mathbf{I}\;.
\end{equation}
In their experimental study, they used $\beta$ to measure 
the local anisotropy (elongation magnitude) of the loop tensors
of individual void cells.
The current study applies the same methods to analyze heterogeneity in
the local fabric tensor.
\par
Satake~\cite{Satake:1982a} proposed the fabric tensor as a measure
of particle arrangement in a granular material, and we
use a local form, $\mathbf{F}^{k}$, to analyze fabric heterogeneity:
\begin{equation}
F_{pq}^{k} = \frac{1}{n^{k}} \sum_{j=1}^{n^{k}} \eta_{p}^{\,j}\eta_{q}^{\,j}\;,
\end{equation}
where the tensor for a particle $k$ involves its $n^{k}$ contacts.
Superscript $j$ denotes the $j$th contact with particle $k$
(Table~\ref{table:superscripts}).
Vectors $\boldsymbol{\eta}^{j}$ are unit vectors in the directions
of the branch vectors that join the center of particle $k$ with the centers of
its contacting neighbors.
The assembly average $\overline{\mathbf{F}}$ is computed from the sum 
of local values for all $N_{\mathrm{eff}}$ particles that
are included in (attached to) the particle graph,
\begin{equation} \label{eq:Fbar}
\overline{\mathbf{F}} = \frac{1}{2 N_{\mathrm{eff}}}
\sum_{k=1}^{N_{\mathrm{eff}}} n^{k} \mathbf{F}^{k} \;.
\end{equation}
Studies have shown that $\overline{\mathbf{F}}$ becomes increasingly
anisotropic during deviatoric loading, with the major
principal direction of $\overline{\mathbf{F}}$ becoming more aligned with the
direction of compressive loading~\cite{Oda:1982a,Thornton:2000a}.
\par
\begin{figure}
  \centering
  \includegraphics{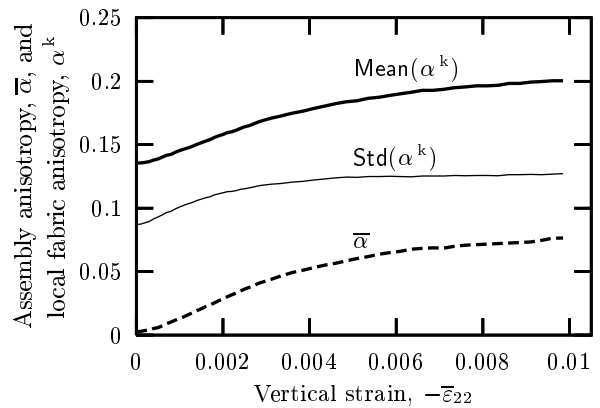}
\caption{Changes in the average and local fabric anisotropies 
during biaxial compression.}
\label{fig:fabric}
\end{figure}

The current study considers variability in the local anisotropy of
fabric.
We apply Eq.~\ref{eq:beta} to the local fabric tensor $\mathbf{F}^{k}$
to compute a local measure $\alpha^{k}$ of fabric anisotropy:
\mbox{$\mathbf{T} \rightarrow \mathbf{F}^{k}$},
\mbox{$\beta \rightarrow \alpha^{k}$}.
Fig.~\ref{fig:fabric} shows the results for the biaxial compression
tests.
The average fabric anisotropy of the entire assembly, $\overline{\alpha}$,
increases with loading (Eqs.~\ref{eq:beta} and~\ref{eq:Fbar}), 
a results that is consistent with previous
experiments.
As would be expected, the mean local anisotropy, $\mathsf{Mean}(\alpha^{k})$,
is larger than the average assembly anisotropy $\overline{\alpha}$,
and the increase in local anisotropy parallels that of the entire assembly.
The results also show, however, that the standard deviation
of fabric anisotropy increases with strain.
The increase in $\mathsf{Std}(\alpha^{k})$
suggests that the geometric arrangement of particles becomes
more varied during loading.
\subsection{Inter-particle movements} \label{sec:move}
The change in stress within a dry granular material is
due to local changes in the inter-particle
forces that result from the relative shifting of particles during 
assembly deformation.
The simplest models of this mechanism are based upon
the interactions of particle pairs that are constrained
to move in accord with a homogeneous deformation field.
Bathurst and Rothenburg~\cite{Bathurst:1988a} studied the
inter-particle movements at small strains in
the biaxial compression of a disk assembly.
Their results demonstrate that, on average, the inter-particle
movements at small strains are less than those that would be consistent with
uniform deformation (see also~\cite{Kruyt:2002a}).
The current study addresses the non-conformity of inter-particle movements
relative to the average deformation,
the diversity of this non-conformity, its evolution during loading,
and the spatial coherence of the non-conformity.
In this regard, we consider only those particles that are
included in the particle graph at a particular
stage of loading.
The relative velocity $\widehat{\mathbf{v}}^{j}$ of two particles
$k_{1}$ and~$k_{2}$ is the difference in their velocities
\begin{equation}
\widehat{\mathbf{v}}^{j} = \mathbf{v}^{k_{2}} -  \mathbf{v}^{k_{1}}\;,
\end{equation}
where index $j$ represents the contacting pair \mbox{$(k_{1},k_{2})$}.
The relative movement that would be consistent with homogeneous deformation
is the product $\overline{\mathbf{L}}\,\mathbf{l}^{j}$, 
where $\overline{\mathbf{L}}$ is the average velocity gradient of the assembly,
and $\mathbf{l}^{j}$ is the branch vector between the 
centers of particles $k_{1}$ and $k_{2}$ (Table~\ref{table:superscripts}).
\par
The quantities in Eqs.~(\ref{eq:parallel}--\ref{eq:circ}) can
be applied to
describe the conformity (or non-conformity) and
diversity of the local, inter-particle
movements $\widehat{\mathbf{v}}^{j}$ with respect to the
mean-field displacement $\overline{\mathbf{L}}\,\mathbf{l}^{j}$.
We begin by considering only pairs of particles that are in
direct contact during biaxial compression
(the number of these pairs ranges from 17,600 to 21,300 for
the 10,816 particles),
although we will consider more distant pairs in a later paragraph.
\par
\begin{figure}
  \centering
  \includegraphics{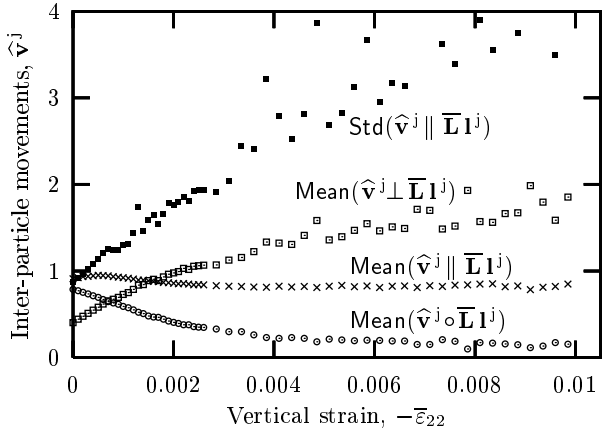}
\caption{Evolution of the non-conformity and heterogeneity
of inter-particle motions $\widehat{\mathbf{v}}^{j}$ 
during biaxial compression.
The motions are for particle pairs $j$ that are in direct contact
($\rho=1$).  Over 18,000 pairs are represented in each point.}
\label{fig:contactMove_strain}
\end{figure}

The evolution of measures~(\ref{eq:parallel}--\ref{eq:circ})
are shown in Fig.~\ref{fig:contactMove_strain}.
The average inter-particle motions 
$\widehat{\mathbf{v}}^{j}$ are consistently less than the
mean-field motions, as is shown by a mean conformity
$\mathsf{Mean}(\widehat{\mathbf{v}}^{j} \!\parallel 
\overline{\mathbf{L}}\,\mathbf{l}^{j})$
less than 1.
This result is consistent with studies~\cite{Kruyt:2002a}
and~\cite{Bathurst:1988a}, 
which investigated the local behavior at small strains.
Figure~\ref{fig:contactMove_strain} shows that the mean
conformity,
$\mathsf{Mean}(\widehat{\mathbf{v}}^{j} \!\!\parallel
\overline{\mathbf{L}}\,\mathbf{l}^{j})$,
is modestly reduced during loading,
from about 0.91 to about 0.82.
As we will see, however, the diversity of the fluctuations
can be quite large.
Both the non-conformity and heterogeneity of inter-particle
motions are indicated
by the additional measures
$\mathsf{Mean}(\widehat{\mathbf{v}}^{j} \!\!\perp 
\overline{\mathbf{L}}\,\mathbf{l}^{j})$
and
$\mathsf{Mean}(\widehat{\mathbf{v}}^{j} \!\circ 
\overline{\mathbf{L}}\,\mathbf{l}^{j})$.
If the local motions were in uniform conformance with the 
assembly deformation,
these two measures would have values of~0 and~1 respectively.
At large strains, the value of 
$\mathsf{Mean}(\widehat{\mathbf{v}}^{j} \!\perp
\overline{\mathbf{L}}\,\mathbf{l}^{j})$
approaches~2, compared with a value of
$\mathsf{Mean}(\widehat{\mathbf{v}}^{j} \!\!\parallel\!
\overline{\mathbf{L}}\,\mathbf{l}^{j})$
of about~0.82.
These results reveal that, on average and at
large strains, the components of inter-particle
movements that are \emph{orthogonal} to their mean-field directions
can be more than twice as large as the components that
are aligned with the mean-field directions
(Eqs.~\ref{eq:parallel} and~\ref{eq:perp}).
This lack of vector alignment is also indicated by the 
cosine-type measure
$\mathsf{Mean}(\widehat{\mathbf{v}}^{j} \circ
\overline{\mathbf{L}}\,\mathbf{l}^{j})$,
which is reduced to a value of about 0.15 (see Eq.~\ref{eq:circ}).
At the end of the test, fully 40\% of inter-particle motions were in
the ``wrong'' direction, with values
$\widehat{\mathbf{v}}^{j}\cdot(\overline{\mathbf{L}}\,\mathbf{l}^{j})<0$.
The fourth measure in Fig.~\ref{fig:contactMove_strain} is
$\mathsf{Std}(\widehat{\mathbf{v}}^{j} \parallel
\overline{\mathbf{L}}\,\mathbf{l}^{j})$,
which displays a rather extreme degree of nonuniformity
in the components of inter-particle movements that are
parallel to the mean-field directions.
This nonuniformity is particularly sizable
at large strains.
A set of random vectors of uniform length would have a value
of 
$\mathsf{Std}(\widehat{\mathbf{v}}^{j} \!\!\parallel\!
\overline{\mathbf{L}}\,\mathbf{l}^{j})$
of only 0.5 (Table~\ref{table:values}), 
a value several times smaller than those 
in Fig.~\ref{fig:contactMove_strain}. 
Such large values
indicate a substantial heterogeneity in both the magnitudes
and directions of the inter-particle
movements $\widehat{\mathbf{v}}^{j}$.
\par
\begin{figure}
  \centering
  \includegraphics{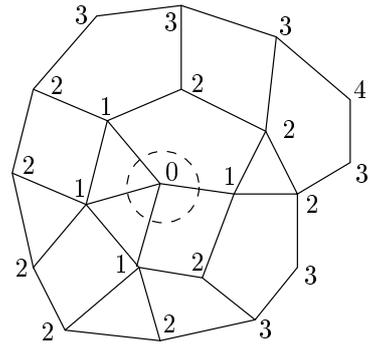}
\caption{Discrete distances $\rho$ from a reference particle 0.
The vertices represent particle centers; 
edges represent particle contacts.}
\label{fig:distance}
\end{figure}

We can also use the biaxial compression simulation to investigate
the spatial correlation of inter-particle movements and the length scale
at which the inter-particle movements approximate the mean deformation
field.
Kruyt and Rothenburg~\cite{Kruyt:2002a} measured the spatial
correlation of movements at small strains by using a 2-point
correlation technique.
In the current study, we do not consider all possible particle
pairs, but instead use only those pairs of particles that are included
in (attached to) the particle graph, as only these particles participate
directly in the deformation and load-bearing mechanisms.
This limitation suggests a \emph{discrete metric} $\rho$
for describing the distance between two particles $k_{1}$ and $k_{2}$.
The distance $\rho(k_{1},k_{2})$ is the least number
of contacts (graph edges) that must be traversed to connect
$k_{1}$ and $k_{2}$ (Fig.~\ref{fig:distance}).
The results in Fig.~\ref{fig:contactMove_strain}, 
which have already been described, were
collected from the sets of all particle pairs at a 
discrete distance of~1,
i.e. the sets $\{ (k_{1},k_{2}\mbox{):}\; \rho(k_{1},k_{2})=1 \}$
at various stages of loading.
The discrete metric does not provide angle or size, so all subsequent
calculations with the objects $\widehat{\mathbf{v}}^{j}$,
$\overline{\mathbf{L}}$, and $\mathbf{l}^{j}$ were, of course,
performed in Euclidean space, but only on the selected particle pairs.
\par
\begin{figure}
  \centering
  \includegraphics{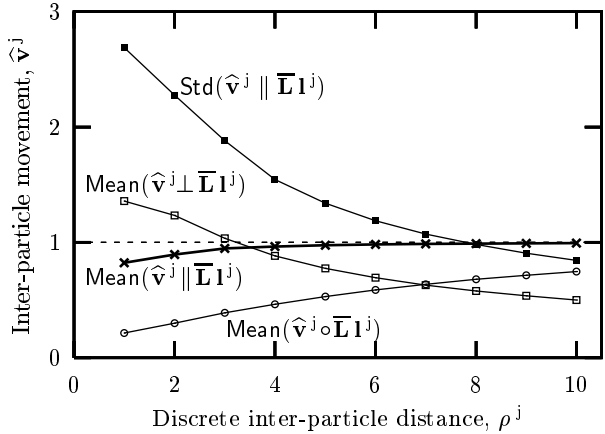}
\caption{The correlation of inter-particle motions with the
discrete distance $\rho$ between particle pairs at
a strain $\overline{\varepsilon}_{22}=-0.005$.
The superscript $j$ represents a pair of particles $(k_{1},k_{2})$
that are separated by distance $\rho$.
The results at $\rho=1$ involve 18,000 pairs; results
at $\rho=10$ involve over 250,000 pairs.}
\label{fig:Contact_move_dist_005}
\end{figure}

Figure~\ref{fig:Contact_move_dist_005} 
shows the non-conformity and heterogeneity of 
inter-particle movements $\widehat{\mathbf{v}}^{j}$
for particle pairs $j$ at distances
$\rho$ of~1 to~10, 
but at the single large strain $\overline{\varepsilon}_{22}=-0.005$
(see Fig.~\ref{fig:crs_q}).
(The results for $\rho=10$ involve over one-quarter million particle pairs.)
As would be expected, the average conformity of 
the observed inter-particle movements
with their corresponding mean-strain movements 
improves with an increasing discrete
distance between the pairs.
This improved conformity is evidenced by increases
in the measures
$\mathsf{Mean}(\widehat{\mathbf{v}}^{j} \!\parallel
\overline{\mathbf{L}}\,\mathbf{l}^{j})$
and
$\mathsf{Mean}(\widehat{\mathbf{v}}^{j} \!\circ
\overline{\mathbf{L}}\,\mathbf{l}^{j})$
and in the reduction of 
$\mathsf{Mean}(\widehat{\mathbf{v}}^{j} \!\perp
\overline{\mathbf{L}}\,\mathbf{l}^{j})$.
However, at a distance of $\rho=10$ and at the strain 
$\overline{\varepsilon}_{22}=-0.005$,
the values of these three measures are about the same as
those at distance $\rho=1$ with zero strain,
$\overline{\varepsilon}_{22}\approx 0$.
That is,
at the large strain of $-0.005$, the non-conformity of motion at a distance of
about~8--10 particle diameters is no better than the
modestly substantial non-conformity of neighboring particles at small
strains.
\par
\begin{figure}
  \centering
  \includegraphics{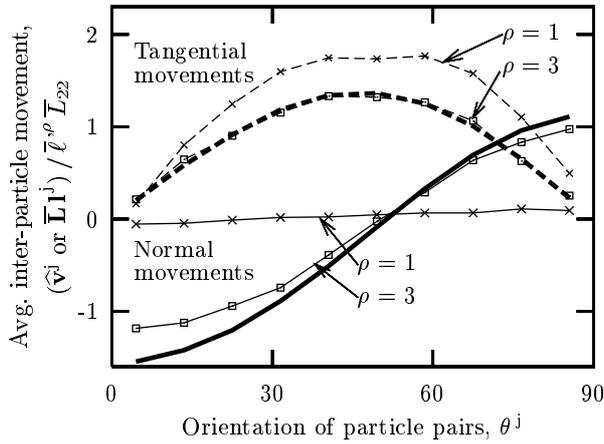}
\caption{The average normal and tangential motions of particle
pairs as a function of the pair orientation $\theta^{j}$.
Mean-field motions $\overline{\mathbf{L}}\,\mathbf{l}^{j}$
are represented by heavy lines; whereas, the averaged
actual inter-particle motions are the lighter lines.
Values are given for pairs having discrete distances
$\rho$ of~1 and~3.  The compressive strain $\overline{\varepsilon}_{22}$
is $-0.005$ (see Fig.~\ref{fig:crs_q}).}
\label{fig:Contact_move_orient_005}
\end{figure}
\begin{figure}
  \centering
  \includegraphics{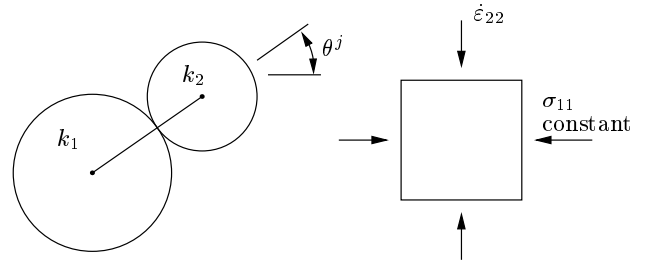}
\caption{Orientation angle $\theta^{j}$ for a particle pair.}
\label{fig:theta}
\end{figure}

The conformity between the actual and mean-field motions is particularly
poor at large strains if we consider only the \emph{normal} motions
between the particle pairs that are in direct contact (i.e. with $\rho=1$).
Figure~\ref{fig:Contact_move_orient_005} shows the assembly averages of the
normal and tangential motions of those particle pairs that are separated
by distances $\rho$ of~1 and~3, at the large strain
$\overline{\varepsilon}_{22}=-0.005$.
These motions are plotted against the orientation
angles $\theta^{j}$ of the pairs (Fig.~\ref{fig:theta}),
and advantage has been taken of the loading symmetry by
folding the angles
$\theta^{j}$ into the single quadrant 0$^{\circ}$ to 90$^{\circ}$.
The normal inter-particle motions are the inner products
\mbox{$\widehat{\mathbf{v}}^{j} \!\cdot\! \boldsymbol{\eta}^{\,j}$}, 
where $\boldsymbol{\eta}^{\,j}$ is the unit vector
aligned with the branch vector $\mathbf{l}^{j}$ that connects the centers
of a particle pair $j = (k_{1},k_{2})$.
Figure~\ref{fig:Contact_move_orient_005} compares the 
averages of these values with the
corresponding averages of the mean-field motions
$\overline{\mathbf{L}}\,\mathbf{l}^{j}$
(the latter are represented with heavy lines).
The results have been normalized by dividing by the average
length $\ell^{,\rho} = \langle |\mathbf{l}^{j,\rho}| \rangle$
for a particular separation $\rho$ and by the strain rate
$\overline{L}_{22}$.
Figure~\ref{fig:Contact_move_orient_005} shows that, at large strains,
the movements of contacting particles ($\rho=1$) are predominantly tangential,
and that the mean normal motion is quite small.
That is, at $\rho = 1$ and at large strains,
the normal inter-particle movements are grossly overestimated
by the mean-field motion $\overline{\mathbf{L}}\,\mathbf{l}^{j}$.
At a distance $\rho=3$, the motions are, on average, in
much closer conformity with those predicted by a mean-field assumption.
The apparent conformity at $\rho=3$ 
in Fig.~\ref{fig:Contact_move_orient_005}
is, however, based upon an average of movements,
and the true diversity in their values is more appropriately reflected in
the measures 
$\mathsf{Mean}(\widehat{\mathbf{v}}^{j} \!\perp
\overline{\mathbf{L}}\,\mathbf{l}^{j})$,
$\mathsf{Mean}(\widehat{\mathbf{v}}^{j} \!\circ
\overline{\mathbf{L}}\,\mathbf{l}^{j})$,
and
$\mathsf{Std}(\widehat{\mathbf{v}}^{j} \!\parallel
\overline{\mathbf{L}}\,\mathbf{l}^{j})$,
which are reported in Figs.~\ref{fig:contactMove_strain} 
and~\ref{fig:Contact_move_dist_005}.
\subsection{Deformation heterogeneity} \label{sec:deform}
\begin{figure}
  \centering
  \includegraphics{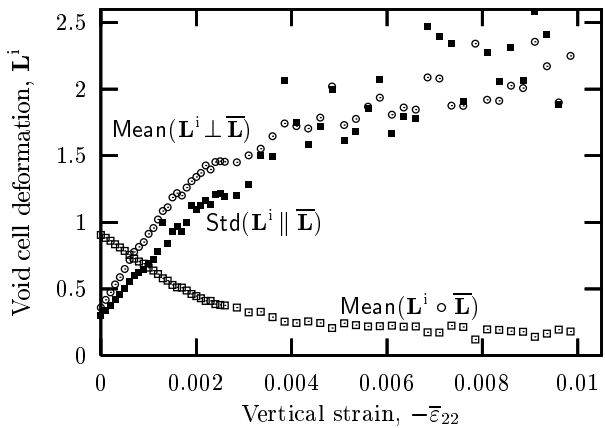}
\caption{The evolution of deformation non-conformity
and heterogeneity during biaxial compression.
Each point represents the deformations $\mathbf{L}^{i}$
in over 7500 void cells,  
where superscript $i$ represents an $i$th void cell.}
\label{fig:def_var_strain}
\end{figure}

Micro-scale deformations within a 2D granular material
can be computed by considering the small polygonal void cells
as being representative micro-regions among particle 
clusters (Fig.~\ref{fig:graph})~\cite{Bagi:1996a,Kruyt:1996a,Kuhn:1999a}.
The region of 10,816 particles can be partitioned into over 7500
of these void cells.
The average velocity gradient 
$\mathbf{L}^{i}$ within a single polygonal void cell $i$ is computed
from the motions of the particles at its vertices.
These local velocity gradients can then be compared with the
average assembly gradient $\overline{\mathbf{L}}$,
and the measures in Eqs.~(\ref{eq:parallel}--\ref{eq:circ})
can be used to investigate the non-conformity and 
heterogeneity of local deformations.
Figure~\ref{fig:def_var_strain} shows the evolution of
these measures in the course of a biaxial compression test.
At small strains, the local deformations are modestly aligned
with the averaged deformation:
the average cosine of alignment, 
$\mathsf{Mean}(\mathbf{L}^{i} \!\circ \overline{\mathbf{L}})$,
is 0.91, only slightly lower than~1,
and the average component of the local gradient
$\mathbf{L}^{i}$ that is perpendicular to the assembly
average $\overline{\mathbf{L}}$ is about 35\% of $|\overline{\mathbf{L}}|$.
At larger strains, the local deformations are, on average, far more deviant
and exhibit a much larger dispersion of values.
The standard deviation of the aligned deformations,
$\mathsf{Mean}(\mathbf{L}^{i} \!\parallel \overline{\mathbf{L}})$,
becomes more than twice its mean value of~1.
Deformations that are orthogonal to $\overline{\mathbf{L}}$
become, on average, much larger than those parallel to 
$\overline{\mathbf{L}}$
(compare the
$\mathsf{Mean}(\mathbf{L}^{i} \!\perp \overline{\mathbf{L}})$
in Fig.~\ref{fig:def_var_strain}
with a
$\mathsf{Mean}(\mathbf{L}^{i} \!\parallel \overline{\mathbf{L}})$
of 1).
\par
\begin{figure}
  \centering
  \parbox{8.5cm}
  {\centering%
  \includegraphics{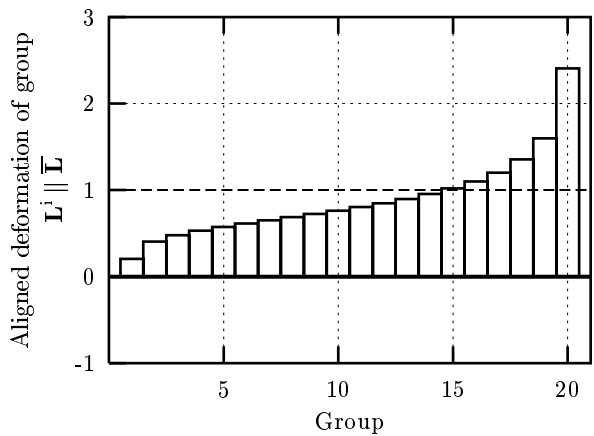}\\[0ex]
  \small{(a) $\overline{\varepsilon}_{22} = -0.0005$}\\[3.0ex]
  \includegraphics{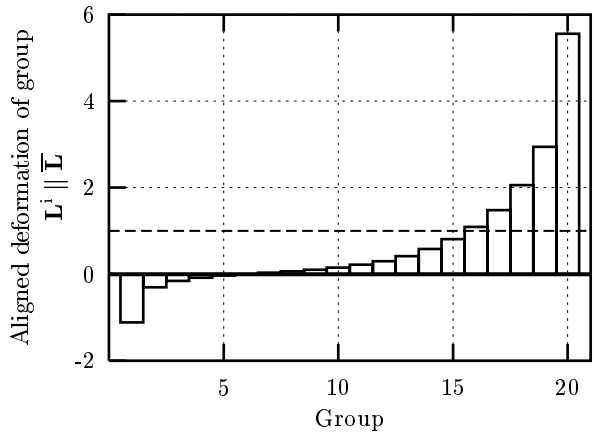}\\[0ex]
  \small{(b) $\overline{\varepsilon}_{22} = -0.005$}
  }
\caption{Distributions of the aligned deformation of void cells
at two strains.  The void cells have been grouped according
to a ranking of their $\mathbf{L}^{i} \!\parallel \overline{\mathbf{L}}$
values
(10,900 and 8300 void cells are included at the two strains).}
\label{fig:Group_align}
\end{figure}
\begin{figure}
  \centering
  \includegraphics{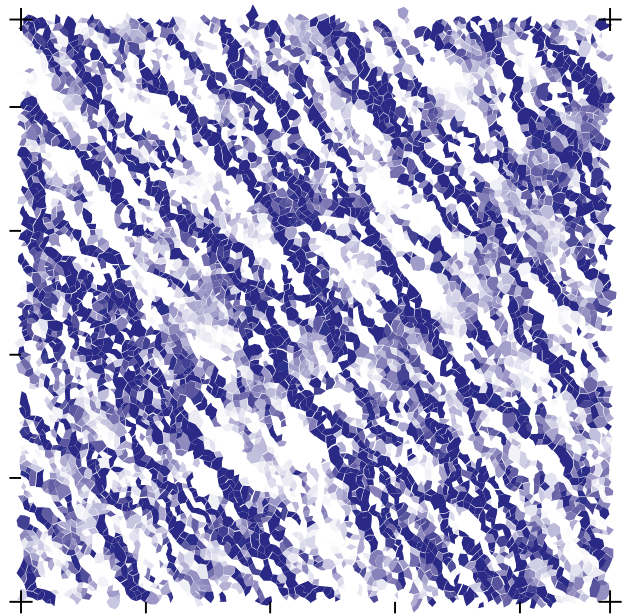}
\caption{The presence of right-shear microbands at
strain $\overline{\varepsilon}_{22} = -0.0005$.
The local void cell deformations $\mathbf{L}^{i}$ have
been filtered as $\mathbf{L}^{i} \boldsymbol{\Phi}$,
where the filter $\boldsymbol{\Phi} = [0.49\;0.41 ; -0.58\; -0.49]$
captures a deformation mode that produces shearing that is
downward and to the right.
A complementary set of left-shear microbands would be present with
the use of an alternative filter.
The gray scale illustrates the magnitudes of
the local filtered deformations, but some of the white regions have
negative filtered values in this monochrome plot.}
\label{fig:microbands}
\end{figure}

This non-conformity and heterogeneity is also illustrated in
Fig.~\ref{fig:Group_align}, 
which shows the distributions of aligned deformations
at moderate and large compressive strains, $\overline{\varepsilon}_{22}$
of $-0.0005$ and $-0.005$.
In each figure, the void cells have been placed into~20 bins,
arranged according to a ranking of the 
aligned deformations
$\mathbf{L}^{i} \!\parallel \overline{\mathbf{L}}$
of each, $i$th void cell.
At moderate strains, the 10\% of most contributory void cells
participate disproportionately in the average assembly
deformation and about 6.5 times more than
the lowest 10\% of void cells (Fig.~\ref{fig:Group_align}a).
At the larger strain of $-0.005$, about 22\% of the material makes
a \emph{negative} contribution to the overall assembly deformation,
and, in a sense, is deforming in the ``wrong'' direction
(Fig.~\ref{fig:Group_align}b).
As another measure of this heterogeneity at large strain, the 31\% of
most contributory void cells could account, by themselves, for
the entire assembly deformation.
This situation is akin to that of a material in which a shear band
has developed, where intense shearing within the band
is accompanied by unloading outside of the band.
No shear bands were observed in the current simulations,
although another type of localization, in the form of multiple
non-persistent \emph{micro-bands}, 
was present throughout the biaxial compression
test.
This type of deformation patterning, described in~\cite{Kuhn:1999a},
was subtly present at the start of deformation and became more
pronounced as deformation proceeded.
Microband localization accounts for much of the deformation
heterogeneity that is recorded in 
Figs.~\ref{fig:def_var_strain} and~\ref{fig:Group_align}.
An example of micro-band patterning at small strain is shown in
Fig.~\ref{fig:microbands}, in which the local, void cell deformations
$\mathbf{L}^{i}$ have been filtered to highlight
a right-shear deformation mode
(see~\cite{Kuhn:1999a} for a discussion of the visualization
technique).
\subsection{Particle rotation heterogeneity} \label{sec:rotate}
\begin{figure}
  \centering
  \includegraphics{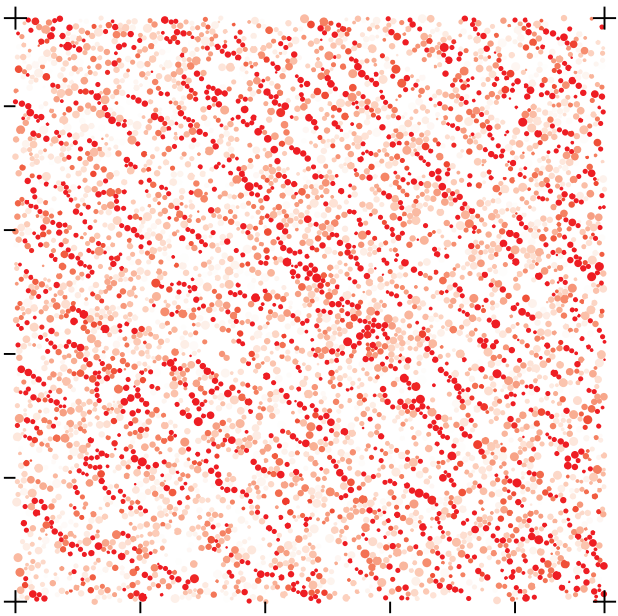}
\caption{Particle spins in a biaxial compression test at
strain $\overline{\varepsilon}_{22} = -0.0005$.  Only 
clockwise spinning particles are shown in the plot.}
\label{fig:rotations}
\end{figure}

Particle rotations in granular materials are known to be large,
particularly in 2D assemblies of circular disks.
Dedecker et~al.~\cite{Dedecker:2000a} found that the standard
deviation of the particle rotation rates could be several times larger
than the average strain rate of an assembly.
Calvetti et~al.~\cite{Calvetti:1997a} reported that the variability
of particle rotations increased consistently with increasing
strain.
Figure~\ref{fig:rotations} shows that this variability 
is expressed in a
spatial patterning of 
particle rotations.
The figure is taken at the
moderate strain $\overline{\varepsilon}_{22}$ of $-0.0005$, but
\emph{only counter-clockwise} rotations are shown in this 
monochrome plot, where the shading depends upon the 
dimensionless rotation rate $\omega^{k}/ |\overline{\mathbf{L}}|$.  
The most rapidly rotating particles are usually aligned in chain-like 
patterns oblique to the principal stress directions.
These chains are closely associated with microbands, 
as can be seen by comparing Figs.~\ref{fig:microbands} 
and~\ref{fig:rotations}~\cite{Kuhn:1999a}.
\subsection{Stress heterogeneity} \label{sec:stress}
The transmission of force among particles occurs in a
non-uniform manner, with certain chains of particles bearing
a disproportionate share of the surface tractions.
These force chains have been widely observed, and several related
references are given in Table~\ref{table:class1}.
The current study concerns the distribution of \emph{stress}
among an assembly's particles.
In two previous studies,
the local variation of stress within stacks of rods
has been studied by withdrawing groups of rods and
measuring the removal force~\cite{Bacconnet:1992a,Auvinet:1992a}.
The DEM simulations of 
the current study allow the direct computation
of stress $\boldsymbol{\sigma}^{k}$ within each, $k$th disk:
\begin{equation}
\sigma_{pq}^{k} = \frac{r^{k}}{A^{k}}\sum_{j=1}^{n^{k}}
\eta_{p}^{\,j} f_{q}^{\,j} \;,
\end{equation}
where summation is over the $n^{k}$ contacts $j$ of the particle $k$,
$r^{k}$ is the disk radius, $\boldsymbol{\eta}^{k}$ is the
unit normal vector, and $\mathbf{f}^{j}$ is the contact force.
Satake~\cite{Satake:1992a} and 
Kruyt and Rothenburg~\cite{Kruyt:2002a} have
described a dual of the particle graph that could be used
to compute a representative particle area $A^{k}$ that includes a portion
of the void space around a particle.
To compute a local stress that can be compared with the
average assembly stress, we instead use the (solid) disk area 
$\pi (r^{k})^{2}$ and
simply divide it by the assembly-average solid fraction.
\par
\begin{figure}
  \centering
  \includegraphics{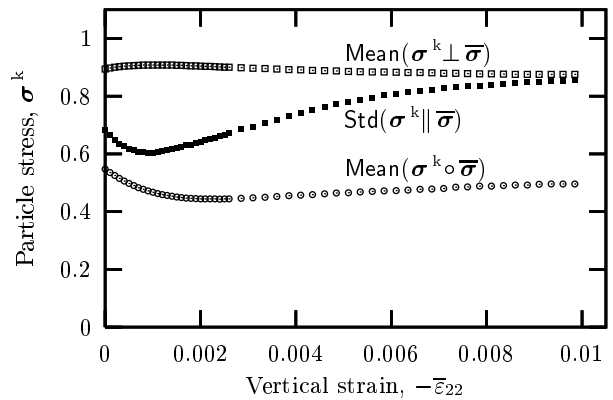}
\caption{The evolution of stress non-conformity and heterogeneity during
biaxial compression.}
\label{ref:stress_var_strain}
\end{figure}

Figure~\ref{ref:stress_var_strain} 
shows the evolution of non-conformity and heterogeneity
in the local stress $\boldsymbol{\sigma}^{k}$
(eqs.~\ref{eq:parallel}--\ref{eq:circ}).
The average, cosine-type alignment of the local stress,
$\mathsf{Mean}(\boldsymbol{\sigma}^{k} \!\circ
\overline{\boldsymbol{\sigma}})$,
is less than 0.6, but there is little
change in this average alignment during loading.
The spatial variation in local stress, as measured by
$\mathsf{Std}(\boldsymbol{\sigma}^{k} \!\parallel 
\overline{\boldsymbol{\sigma}})$,
decreases at small strains, but then increases at larger strains.
At large strains,
all three measures in Fig.~\ref{ref:stress_var_strain} 
depict a greater conformity
and homogeneity of stress than was found with inter-particle movements
and void cell deformations
(\textit{cf} Figs.~\ref{fig:contactMove_strain},
\ref{fig:def_var_strain} and~\ref{ref:stress_var_strain}).
This greater regularity is likely due to the stress being represented
in its status, whereas movement and deformation were represented
in their rates.
At small strains, however, the three measures 
in Fig.~\ref{ref:stress_var_strain} show less conformity and heterogeneity
in stress than in the inter-particle movements.
The diversity of stress at small strain is primarily
the inheritance of the initial particle packing, and this diversity
increases only modestly during loading.
\par
\begin{figure}
  \centering
  \parbox{8.5cm}
  {\centering%
  \includegraphics{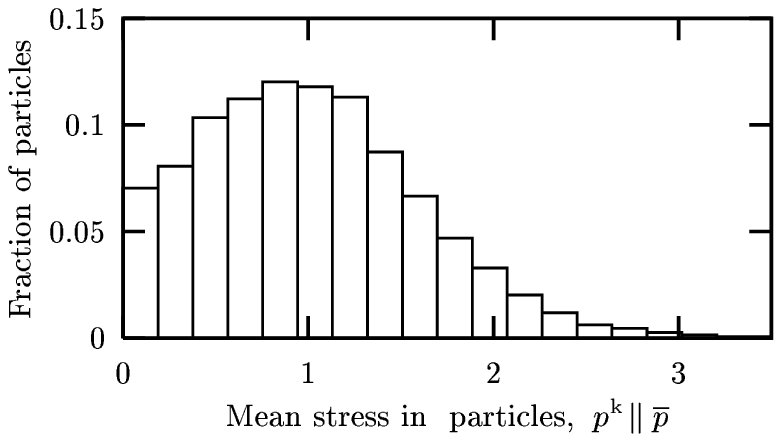}\\[0ex]
  \small{(a) Local mean stress}\\[3.0ex]
  \includegraphics{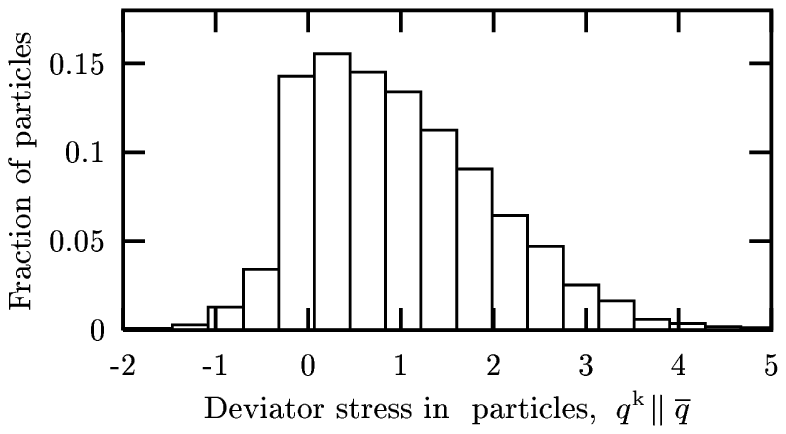}\\[0ex]
  \small{(b) Local deviator stress}
  }
\caption{Participation of the local stress in the average assembly
stress.
Figures~\ref{fig:stress_hist}a and~\ref{fig:stress_hist}b are
histograms of the local participation in the mean and deviator
stresses.
Both figures are compiled from the stresses in over 10,000
particles at the large strain $\overline{\varepsilon}_{22} = -0.005$.}
\label{fig:stress_hist}
\end{figure}

The variation in stress is greatest in its deviatoric component.
Figures~\ref{fig:stress_hist}a and~\ref{fig:stress_hist}b are histograms
of the local mean stress and deviator stress,
defined for particle $k$ as $p^{k}=(\sigma_{11}^{k}+\sigma_{22})/2$
and $q^{k}=(\sigma_{22}^{k}-\sigma_{11})/2$
respectively.
The figure gives these components at the large strain 
$\overline{\varepsilon}_{22} = -0.005$.
Because only compressive force can be delivered between particles,
the local mean stress is uniformly positive, but the
standard deviation of the local mean stress $p^{k}$
is about 0.60 (Fig.~\ref{fig:stress_hist}a).
The standard deviation of the local deviator stress $q^{k}$ is 1.0
(Fig.~\ref{fig:stress_hist}b).
About~15\% of particles have a negative alignment of the
deviator stress, $q^{k} \!\parallel\! \overline{q}$,
and these particles provide a negative contribution toward bearing the
average assembly deviator stress.
\section{Conclusion}
In the paper, we have considered several categories
of heterogeneity in granular materials:
topologic, geometric, kinematic, and static.
In all respects, the heterogeneity can be described, at a minimum,
as being moderate.
Heterogeneity increases during biaxial compressive loading.
In the case of inter-particle movements, the non-uniformity
becomes extreme, and particle motions are only coarsely aligned
with the mean-field movement.
At large strains, significant fluctuations from the mean-field
motion extend to distances of at least eight particle diameters.
Non-uniform motion is expressed in the patterning of local
movements, which includes microband patterning and
rotation chain patterning.
The extent and magnitude of the heterogeneity and its patterning
proffer an imposing challenge to the continuum
representation of granular materials at micro and macro scales,
especially at large strains.
Before such efforts can be productive,
further statistical analyses should be undertaken to
further characterize heterogeneity,
to determine characteristic lengths at which heterogeneity 
dominates the meso-scale behavior, to quantify the heterogeneity
in the local stress rates, and to establish the relationships among
topologic, geometric, kinematic, and static heterogeneities.
%
%
% BibTeX users please use
\bibliographystyle{unsrt}
%
% My bibliographic data base file:  Kuhn.bib
%\bibliography{Kuhn}
% end of file Kuhn.tex

\end{document}